\documentclass[PP]{AEA}
\usepackage{graphicx}
\usepackage{mathptmx}
\usepackage{lipsum}
\usepackage{booktabs}
\usepackage{colortbl}
\usepackage{xcolor}
\usepackage{listings}
\definecolor{lightred}{RGB}{255,230,230}
\definecolor{mediumred}{RGB}{255,150,150}
\definecolor{darkred}{RGB}{255,70,70}
\definecolor{lightblue}{RGB}{230,230,255}
\definecolor{mediumblue}{RGB}{150,150,255}
\definecolor{darkblue}{RGB}{70,70,255}

\newcommand{\zscore}[1]{%
    \ifdim#1pt>1.5pt
        \cellcolor{darkred}#1%
    \else\ifdim#1pt>0.75pt
        \cellcolor{mediumred}#1%
    \else\ifdim#1pt>0pt
        \cellcolor{lightred}#1%
    \else\ifdim#1pt<-1.5pt
        \cellcolor{darkblue}#1%
    \else\ifdim#1pt<-0.75pt
        \cellcolor{mediumblue}#1%
    \else\ifdim#1pt<0pt
        \cellcolor{lightblue}#1%
    \else
        #1%
    \fi\fi\fi\fi\fi\fi
}

\usepackage{natbib}

\draftSpacing{1.5}

\begin{document}

\title{Anonymous Attention and Abuse}
\shortTitle{Anonymous Attention and Abuse}
\author{Florian Ederer, Paul Goldsmith-Pinkham and Kyle Jensen\thanks{Ederer: Boston University, CEPR, ECGI and NBER, ederer@bu.edu. Goldsmith-Pinkham: Yale University and NBER, paul.goldsmith-pinkham@yale.edu. Jensen: Yale University, kyle.jensen@yale.edu. We thank Martin Oehmke and several economists who wish to remain anonymous for helpful comments. The analysis described in this manuscript was allowed to proceed by the Yale HRPP, IRB protocol ID 2000034072.}}
\date{\today}
\pubYear{2025}
\pubMonth{May}
\pubVolume{115}
\pubIssue{1}
\maketitle

Economics Job Market Rumors (henceforth EJMR) occupies a controversial position within the economics profession. While it began as an anonymous forum primarily focused on the annual job market for economics PhDs, its scope has expanded significantly over the years. Today, EJMR serves as both a clearinghouse for job market information and a breeding ground for discussions that range from professional commentary to abusive rhetoric. The forum is infamous for hosting content that is defamatory, misogynistic, and otherwise toxic \citep{wu2018gendered,wu2020gender,ederer2024anonymity}. Its anonymity fosters candid discussions, but it also enables behaviors that contribute to its notoriety. This duality of EJMR---as both a valuable resource and a harmful platform---raises important questions about its role in the economics community.

The influence of the platform is considerable. In early 2023, SimilarWeb estimated that EJMR received 2.5 million monthly visits, surpassing even prominent academic resources such as the National Bureau of Economic Research (NBER) and the American Economic Association (AEA) websites in terms of page views per visit. Such metrics underscore its popularity, but they also highlight its unique position as a digital forum that blends professional and informal exchanges. Despite its popularity, EJMR has been criticized as a ``cesspool of misogyny'' (David Romer quoted in \cite{wolfers2017evidence}), ``4chan for economists'' \citep{lowrey2022sexism}, and evidence of a toxic culture that marginalizes women and underrepresented groups within the profession \citep{wolfers2017evidence}.

\begin{figure}[t]
\label{fig:stack-plot}
\includegraphics[width=\linewidth]{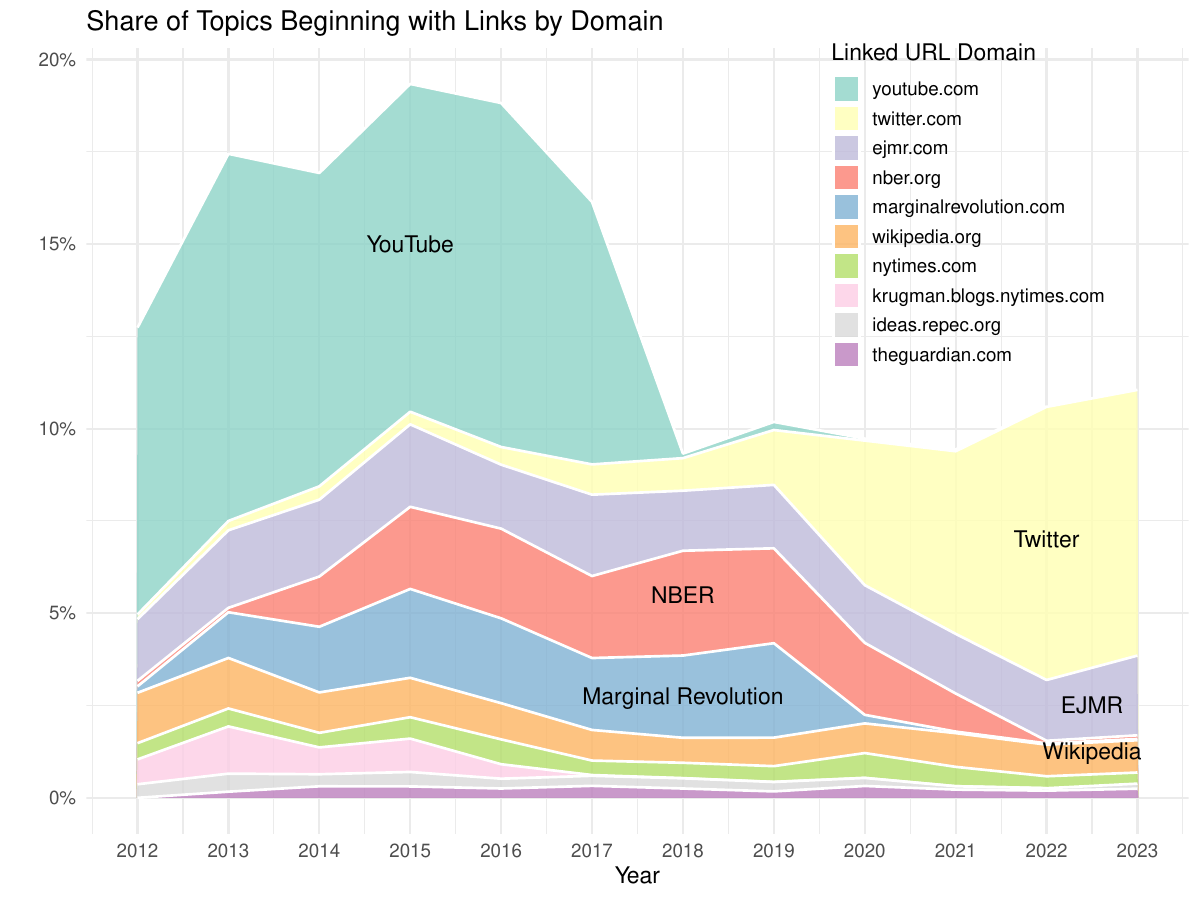}
\caption{Share of discussion topics with links in the first post over time.}
\begin{figurenotes}
This figure plots the share of EJMR topics that begin with a link for the top 10 most popular domains linked on the site over our sample. 
\end{figurenotes}
\end{figure}

\begin{figure}[t]
\label{fig:stack-plot-num}
\includegraphics[width=\linewidth]{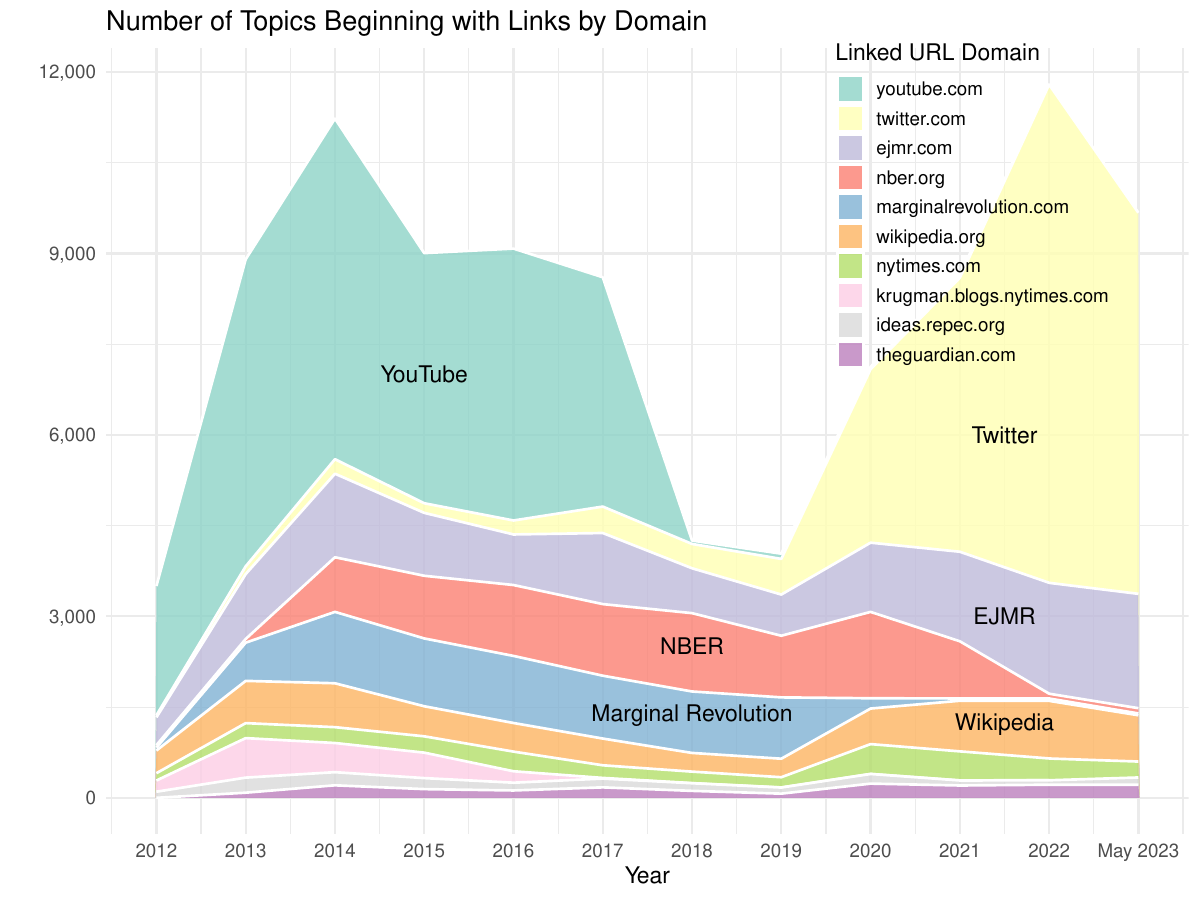}
\caption{Number of discussion topics beginning with links over time.}
\begin{figurenotes}
This figure plots the number of EJMR topics that begin with a link for the top 10 most popular domains linked on the site over our sample.
\end{figurenotes}
\end{figure}



A significant focus of EJMR is its function as an aggregator of external information. In this article we analyze EJMR content from January 2012 to May 2023\footnote{The end of our sample coincides with the public announcement of the results of \cite{ederer2024anonymity} which significantly changed the content and discourse on EJMR.} and show that a sizable proportion of EJMR discussion topics begin with links to other websites, ranging from academic papers and news articles to social media posts and blog entries. Over time, the platform's reliance on external domains has evolved, reflecting broader changes in how economists interact with information. We show that starting in 2018, EJMR saw an explosion in discussions initiated by references to Twitter posts. This shift mirrors Twitter's growing importance as a real-time source of information and debate in academic and public policy circles. The analysis of these linked domains offers insight into how the economics profession engages with a diverse array of online platforms and information sources.

EJMR's transformation also underscores the changing landscape of digital discourse in economics. Initially dominated by academic content, the forum has increasingly incorporated references to social media, highlighting its role as a nexus between professional and public discussions. The platform's toxicity, however, raises ethical questions about its impact on the profession. The prevalence of abusive content, as documented in prior research, contributes to a hostile environment that discourages participation from marginalized groups. This dual nature, as both a valuable resource and a source of harm, necessitates a nuanced examination of its role in shaping economic discourse.

Our paper complements previous findings of \cite{ederer2024anonymity} by analyzing EJMR's evolving interactions with external information sources. We focus on three key aspects: (1) the prevalence and impact of links to external domains; (2) the surge in discussions driven by Twitter posts since 2018; and (3) the categorization of individuals whose tweets and content are discussed on EJMR. Using data on linked domains, we examine how these trends reflect broader changes in the economics profession's digital footprint. Our analysis not only sheds light on EJMR's informational role but also raises critical questions about the forum's implications for inclusivity and professional ethics.

\section{Links to External Domains}

Our examination of EJMR discussions reveals that a significant proportion of discussion topics begin with links to external sites. To measure links to external sites, we use a database of all posts on EJMR from 2012 to May 2023, and identify all links---``anchor'' or ``\texttt{a}'' tags with \texttt{href} atributes in the HyperText Markup Language (HTML). We then roll-up these links to their canonical domain name. For example, we combine the two links \texttt{m.twitter.com} and \texttt{www.twitter.com} into the Twitter domain. We focus on the presence of links that are in the first post of a given topic (a thread of posts) in order to avoid double counting and assess the impact of links driving the conversation on EJMR.

From January 2012 to May 2023, the proportion of topics containing a link in the first post varies between roughly 10 and 20 percent. These links predominantly direct to academic resources, news outlets, and increasingly over time, to the social media platform Twitter, reflecting the shifting priorities and preferences of EJMR users. Early years exhibit a strong preference for academic and blog-based content, such as NBER papers and Marginal Revolution (a popular economics blog) posts. From 2018 onward, Twitter's rapid growth reshaped the forum's linking habits, reflecting broader changes in how economists interact with public and semi-public discussions. This evolution suggests that EJMR has become a microcosm of the economics profession's digital transformation, with implications for how economic ideas are shared and debated.

Figure 1 shows the shares of topics with links in the first post by the domain to which the links point. At the beginning of our sample, approximately 10 percent of the opening posts contain links that direct to YouTube. However, beginning in 2018 the share of topics linking to YouTube disappears almost entirely due to automatic moderation instituted on EJMR. This automatic moderation substantially reduced the ability to post links to YouTube as such posts increasingly contained repetitive spam messages. 

\begin{table}
\footnotesize
  \makebox[\linewidth][c]{
\begin{tabular}{llllllr}
\toprule
& & \multicolumn{4}{c}{Z-score of average post classification} & Number of\\
\cmidrule{3-6}
Twitter Username & LLM Description & Hate Speech & Negativity & Misogynistic & Toxic & Topics\\
\midrule
\texttt{realChrisBrunet} & Right-wing commentator and controversial social media figure & \zscore{-0.127} & \zscore{0.405} & \zscore{0.069} & \zscore{-0.62} & 325\\
\texttt{elonmusk} & Billionaire entrepreneur and technology influencer & \zscore{-0.332} & \zscore{0.172} & \zscore{-0.146} & \zscore{0.056} & 302\\
\texttt{text35237388338} & N/A - Twitter account has been deleted/unavailable & \zscore{-0.37} & \zscore{-7.121} & \zscore{-0.335} & \zscore{-0.851} & 214\\
\texttt{RichardHanania} & Conservative commentator and president of policy center & \zscore{0.12} & \zscore{-0.531} & \zscore{-0.029} & \zscore{1.04} & 189\\
\texttt{Claudia\_Sahm} & Economist and expert on fiscal policy and Fed & \zscore{-0.216} & \zscore{0.187} & \zscore{1.974} & \zscore{0.116} & 172\\
\addlinespace
\texttt{jenniferdoleac} & Economist and criminal justice policy expert & \zscore{-0.003} & \zscore{-0.131} & \zscore{2.598} & \zscore{0.657} & 162\\
\texttt{Noahpinion} & Economics blogger and commentator with a humorous style & \zscore{-0.08} & \zscore{0.065} & \zscore{0.026} & \zscore{0.223} & 158\\
\texttt{JustinWolfers} & Economics professor and commentator on public policy & \zscore{-0.282} & \zscore{0.273} & \zscore{-0.335} & \zscore{0.048} & 132\\
\texttt{libsoftiktok} & Conservative social media commentator and agitator & \zscore{-0.264} & \zscore{0.445} & \zscore{0.191} & \zscore{0.232} & 105\\
Black Female Economist* & N/A - Twitter account has been deleted/unavailable & \zscore{-0.144} & \zscore{0.301} & \zscore{-0.335} & \zscore{-0.205} & 105\\
\addlinespace
\texttt{visegrad24} & Conservative news aggregator and commentator on current affairs & \zscore{0.225} & \zscore{-0.139} & \zscore{-0.335} & \zscore{-0.306} & 103\\
\texttt{realDonaldTrump} & Former U.S. President and conservative political figure & \zscore{-0.136} & \zscore{-0.691} & \zscore{-0.335} & \zscore{0.215} & 102\\
\texttt{text3863} & Christian evangelist and religious commentator & \zscore{-0.37} & \zscore{-6.474} & \zscore{-0.335} & \zscore{-0.851} & 99\\
\texttt{disclosetv} & Sensationalist news aggregator with a conservative slant & \zscore{-0.126} & \zscore{0.234} & \zscore{0.274} & \zscore{-0.294} & 92\\
\texttt{nntaleb} & Philosopher and author focused on probability and society & \zscore{-0.123} & \zscore{0.224} & \zscore{0.281} & \zscore{0.276} & 91\\
\addlinespace
\texttt{WallStreetSilv} & Financial market enthusiast and commentator & \zscore{-0.08} & \zscore{1.022} & \zscore{-0.335} & \zscore{-0.355} & 88\\
\texttt{paulkrugman} & Nobel prize winning economist and columnist & \zscore{-0.37} & \zscore{0.106} & \zscore{-0.335} & \zscore{0.102} & 86\\
\texttt{mattyglesias} & Political commentator and journalist & \zscore{0.026} & \zscore{-0.171} & \zscore{0.326} & \zscore{0.358} & 84\\
\texttt{Steve\_Sailer} & Conservative commentator and social critic & \zscore{-0.046} & \zscore{-0.063} & \zscore{0.474} & \zscore{-0.111} & 76\\
\texttt{zerohedge} & Financial and political news and commentary blog & \zscore{-0.37} & \zscore{0.101} & \zscore{-0.335} & \zscore{-0.132} & 75\\
\addlinespace
\texttt{realchrisrufo} & Conservative commentator and cultural critic & \zscore{-0.213} & \zscore{-0.028} & \zscore{0.45} & \zscore{0.047} & 74\\
\texttt{abdcerian} & Insufficient information to classify account owner & \zscore{-0.37} & \zscore{0.979} & \zscore{-0.335} & \zscore{-0.851} & 65\\
\texttt{nypost} & Conservative-leaning news outlet and media organization & \zscore{-0.37} & \zscore{0.848} & \zscore{-0.335} & \zscore{-0.444} & 62\\
\texttt{nexta\_tv} & Eastern European news and media organization & \zscore{0.19} & \zscore{0.368} & \zscore{0.6} & \zscore{-0.21} & 61\\
\texttt{parsel14} & N/A - Twitter account has been deleted/unavailable & \zscore{-0.165} & \zscore{0.634} & \zscore{-0.335} & \zscore{0.786} & 61\\
\addlinespace
\texttt{MrAndyNgo} & Conservative journalist and political commentator. & \zscore{-0.179} & \zscore{0.827} & \zscore{-0.335} & \zscore{0.236} & 60\\
\texttt{RWApodcast} & Pro-Russian geopolitical commentator and podcast host & \zscore{0.081} & \zscore{0.02} & \zscore{-0.335} & \zscore{-0.334} & 60\\
Black Male Economist* & Economics professor & \zscore{-0.169} & \zscore{-0.686} & \zscore{0.669} & \zscore{0.067} & 60\\
\texttt{nytimes} & Mainstream media news organization and publisher & \zscore{-0.176} & \zscore{-0.3} & \zscore{-0.335} & \zscore{0.477} & 57\\
\texttt{JackPosobiec} & Conservative commentator and political influencer & \zscore{0.307} & \zscore{0.394} & \zscore{-0.335} & \zscore{-0.076} & 53\\
\addlinespace
\texttt{DrEricDing} & Public health expert and social media commentator & \zscore{1.014} & \zscore{0.951} & \zscore{-0.335} & \zscore{0.731} & 49\\
\texttt{AlexBerenson} & Conservative commentator and critic of Covid policies & \zscore{-0.37} & \zscore{-0.171} & \zscore{-0.335} & \zscore{-0.549} & 46\\
\texttt{Breaking911} & Alternative news source focusing on sensational headlines & \zscore{0.122} & \zscore{-0.693} & \zscore{-0.335} & \zscore{-0.287} & 46\\
\texttt{elben} & Assistant Professor of Economics specializing in health policy & \zscore{-0.37} & \zscore{0.501} & \zscore{0.956} & \zscore{-0.26} & 44\\
\texttt{NateSilver538} & Statistician and author with political and sports interests & \zscore{-0.084} & \zscore{0.433} & \zscore{-0.335} & \zscore{-0.198} & 43\\
\addlinespace
\texttt{arindube} & Economics professor focusing on labor and inequality & \zscore{-0.099} & \zscore{0.244} & \zscore{-0.335} & \zscore{1.009} & 43\\
\texttt{EJMR\_news} & Anonymous economics forum content aggregator and commentator & \zscore{-0.37} & \zscore{1.142} & \zscore{-0.335} & \zscore{-0.851} & 41\\
\texttt{stillgray} & Conservative commentator and podcast co-host & \zscore{-0.37} & \zscore{0.9} & \zscore{-0.335} & \zscore{0.154} & 41\\
\texttt{DonnieDarkened} & Christian commentator focused on end-times prophecy & \zscore{-0.06} & \zscore{0.886} & \zscore{-0.335} & \zscore{-0.142} & 40\\
\texttt{florianederer} & Economics professor with interest in antitrust and policy & \zscore{-0.37} & \zscore{0.326} & \zscore{-0.335} & \zscore{-0.099} & 40\\
\addlinespace
\texttt{ProfEmilyOster} & Economist and author focused on parenting and health & \zscore{-0.37} & \zscore{0.433} & \zscore{-0.335} & \zscore{-0.851} & 39\\
\texttt{econjobrumors} & Controversial economics commentator and online provocateur & \zscore{-0.37} & \zscore{0.287} & \zscore{-0.335} & \zscore{-0.851} & 39\\
\texttt{haralduhlig} & Economics professor and commentator on politics and academia & \zscore{-0.37} & \zscore{-0.139} & \zscore{-0.335} & \zscore{-0.142} & 39\\
\texttt{subgirl0831} & Adult content creator and social media personality & \zscore{-0.37} & \zscore{1.142} & \zscore{-0.335} & \zscore{-0.076} & 39\\
\texttt{wesyang} & Conservative commentator and writer on social issues & \zscore{2.268} & \zscore{-0.555} & \zscore{-0.335} & \zscore{0.489} & 39\\
\addlinespace
\texttt{alexkokcharov} & N/A - Twitter account has been deleted/unavailable & \zscore{-0.077} & \zscore{-1.041} & \zscore{-0.335} & \zscore{-0.851} & 39\\
\texttt{aylamao3} & N/A - Twitter account has been deleted/unavailable & \zscore{-0.37} & \zscore{1.142} & \zscore{-0.335} & \zscore{-0.851} & 38\\
\texttt{hashtag} & N/A - Twitter account has been deleted/unavailable & \zscore{-0.041} & \zscore{-0.761} & \zscore{1.309} & \zscore{-0.475} & 37\\
\texttt{greg\_price11} & Conservative political commentator and Trump supporter & \zscore{-0.051} & \zscore{0.35} & \zscore{-0.335} & \zscore{-0.486} & 36\\
\texttt{wwwojtekk} & Economist and editor with skeptical commentary & \zscore{-0.37} & \zscore{-1.024} & \zscore{-0.335} & \zscore{0.003} & 36\\
\bottomrule
\end{tabular}}
\caption{Top 50 Twitter accounts mentioned on EJMR}
\begin{tablenotes}
This table shows the Twitter accounts most frequently mentioned on EJMR along with an LLM description (openAI 4o), z-scores of the average EJMR post classification measures of Hate Speech, Negativity, Misogynistic, and Toxic from \citet{ederer2024anonymity}. These measures are standardized into z-scores using the average measures for all Twitter accounts mentioned on EJMR. A z-score equal to 1 reflects a one-standard deviation above the average relative to all linked Twitter users. Red indicates a high z-score and blue a low z-score. Entries marked by * have been anonymized according to the wishes of the account holders. The table shows that the most frequently mentioned Twitter accounts fall into three main groups: economists, right-wing commentators, and journalists. Discussions mentioning prominent female economists on Twitter have high levels of misogyny, in line with the findings of \cite{wu2018gendered,wu2020gender}.
\end{tablenotes}
\end{table}

In the middle of our sample from 2018 to 2020, about 5 percent of topics link to a combination of websites on economic research such as the blog Marginal Revolution, the National Bureau of Economic Research (NBER), and, to a much smaller extent, the economics working paper repository RePEc. The remaining 5 percent of topics link to Twitter, Wikipedia, the New York Times (with a large share early on linking to Paul Krugman's contributions), the Guardian, and other topics on EJMR. Links to blog posts on Marginal Revolution are overwhelmingly posted by a specifically designed Marginal Revolution bot and contain a short summary of the corresponding blog post. Links to the NBER website are mostly to new NBER working papers (96.3\%) and to a much lesser degree to NBER conference schedules and announcements. Whereas the links to working papers are mostly posted by a dedicated EJMR bot, the topics with links to NBER conference schedules and announcements are posted by human EJMR users. 

Since 2018, EJMR has experienced a substantial increase in topics initiated by references to Twitter posts. This shift corresponds to the broader integration of social media, in particular Twitter, within the academic economics ecosystem commonly referred to as \#EconTwitter. The proportion of EJMR topics citing Twitter surged from around 1\% in 2018 to nearly 8\% by 2023. Twitter's role as a disseminator of real-time news and economic research has made it a focal point for EJMR users who engage with its content. At the same time, the proportion (and also the overall number) of the topics linking to Marginal Revolution and to the NBER falls precipitously after 2020 and 2021, respectively. These two domains are entirely displaced by a large increase in the share of topics linking to Twitter and a much smaller increase in the share of topics linking to other topics on EJMR. 

The increasing importance of Twitter (and \#EconTwitter in particular) on EJMR is not just a matter of composition. Figure 2 shows that by 2022 the large increase of activity on EJMR documented by \cite{ederer2024anonymity} and the concurrent rise of Twitter resulted in almost 8,000 EJMR topics with an opening post linking to Twitter, further highlighting Twitter's role as an influential platform for economic discourse.

\section{Referenced Twitter Accounts on EJMR}

Given the rise of importance of Twitter on EJMR it is natural to ask which Twitter accounts receive the most attention on EJMR. Table 1 reports the usernames of Twitter users that are most frequently linked to on EJMR alongside the number of topics which have as an opening post a link to their corresponding Twitter handle. For each of these accounts (except for those wishing to remain anonymous) we retrieved the account's name and description along with the tweets displayed on the first page of the account's profile on Twitter. These tweets were then fed to OpenAI's 4o LLM \citep{achiam2023gpt} for labeling using a brief prompt, listed in Table 2, asking for a five to ten word classification of the Twitter account owner. The resulting labels (e.g, ``Sensationalist news aggregator with a conservative slant'') are also shown in Table 1. 

\begin{table}
\begin{tabular}{|l|}
\hline
\\
\lstset{
basicstyle=\footnotesize, 
stringstyle=\ttfamily, 
showstringspaces=false} 
\begin{lstlisting}[fontadjust]
Here is some information about the "{{user.name}}" @{{user.username}} account on 
Twitter. The account description is as below:

    ```
    {{ user.description }}
    ```

Here are the most recent posts from this account.

    ```
    {% for tweet in tweets %}
    {{ tweet.text }}

    {% endfor %}
    ```

Based on the posts and the description, give me a five to ten word classification of 
this account owner, e.g. ``liberal commentator and agitator'', ``economics professor 
at a public university'', ``german technology magazine'', ``journalist in mainstream 
media'', etc.

Return only the classification, no commentary. Do not be overly specific, for example, 
prefer ``Nobel prize winning economist and columnist'' to ``Nobel prize winner Paul 
Krugman and author of The Conscience''.
\end{lstlisting}
\\
\\
\hline
\end{tabular}
\caption{Prompt for LLM Classification (OpenAi 4o)}
\end{table}

These accounts can be broadly categorized into three main groups: economists, right-wing commentators, and journalists. The group of economists (e.g., \texttt{Claudia\_Sahm}, \texttt{jenniferdoleac}, and \texttt{JustinWolfers}) includes academic and professional economists from diverse institutions whose tweets often serve as springboards for debates on research findings, policy implications, and professional conduct. The second group includes polarizing and predominantly conservative commentators and agitators (e.g., \texttt{realChrisBrunet}, \texttt{RichardHanania}, and \texttt{libsoftiktok}) and reflects EJMR's right-wing slant and engagement with contentious political and social issues. The third group is a collection of news sources and journalistic accounts, many of which have a conservative slant (e.g., \texttt{visegrad24}, \texttt{disclosetv}, and \texttt{nypost}). 

Table 1 further includes z-scores of the average post classification measures of Hate Speech, Negativity, Misogynistic, and Toxic from \citet{ederer2024anonymity} for the opening post of EJMR topics linking to the respective Twitter accounts. These measures are standardized into z-scores such that a z-score equal to 1 reflects a one-standard deviation above the average relative to all linked Twitter users. Positive values, highlighted in red, indicate above-average measures, whereas negative values, shaded in blue, represent below-average measures. For example, the relatively high z-score of 1.014 for Hate Speech for \texttt{DrEricDing} means that EJMR topics linking to this Twitter account contain more hate speech.\footnote{For the Hate Speech and Misogynistic z-scores, there are many accounts with entries of -0.37 and -0.335, respectively. These specific z-scores are the result of zeroes in the raw data.}

Among the 10 most frequently mentioned Twitter accounts there are four economists, including three female economists. EJMR posts referencing two of these female economists (\texttt{Claudia\_Sahm} and \texttt{jenniferdoleac}) have very high average z-scores of 1.974 and 2.598 for the Misogynistic classifier, indicating that EJMR posters discuss them in strongly misogynistic terms compared to all other Twitter accounts mentioned on EJMR. These results are particularly stark examples of the findings of \cite{wu2018gendered,wu2020gender} who also documents the use of gendered and sexist language by EJMR posters discussing female economists. The only other large average z-score for the Misogynistic measure is for EJMR posts referencing \texttt{elben} (z-score Misogynistic = 0.956), an academic economist who has championed LGBTQ-inclusive policies in the economics profession. Although the results for the most frequently mentioned female economists are particularly pronounced, several other prominent economists are also discussed on EJMR in negative or toxic terms such as \texttt{ProfEmilyOster} (z-score Negativity = 0.433) or \texttt{arindube} (z-score Toxic = 1.009).\footnote{The two Twitter accounts (\texttt{text35237388338} and \texttt{text3863}) with exceedingly high positive sentiment (i.e., large negative z-scores for Negativity) are referenced by repetitive spam posts on EJMR containing religious (Christian) messages of love and forgiveness.}

\section{Conclusion}

EJMR's trajectory, marked by increased linkage to social media and broadening external references, underscores the forum's dual role as an information aggregator and a mirror of professional economics. However, its toxic culture raises concerns about inclusivity and the amplification of divisive narratives. Addressing these challenges requires a concerted effort to balance the platform's informational utility with its broader impact on the economics profession.

\bibliographystyle{aea}
\bibliography{references}


\end{document}